\documentclass[11pt]{article}%
\usepackage{amsmath}
\usepackage{graphicx}%
\usepackage{amsfonts}%
\usepackage{amssymb}
\newcommand{\eqnb}{\begin{equation}}
\newcommand{\eqne}{\end{equation}}

\newtheorem{Rem}{Remark}

\begin{document}

\title{\textbf{Super-Exponential Solution in Markovian Supermarket Models: Framework
and Challenge}}
\author{Quan-Lin Li\\School of Economics and Management Sciences \\Yanshan University, Qinhuangdao 066004, China}
\date{January 28, 2011}
\maketitle

\begin{abstract}
Marcel F. Neuts opened a key door in numerical computation of stochastic
models by means of phase-type (PH) distributions and Markovian arrival
processes (MAPs). To celebrate his 75th birthday, this paper reports a more
general framework of Markovian supermarket models, including a system of
differential equations for the fraction measure and a system of nonlinear
equations for the fixed point. To understand this framework heuristically,
this paper gives a detailed analysis for three important supermarket examples:
M/G/1 type, GI/M/1 type and multiple choices, explains how to derive the
system of differential equations by means of density-dependent jump Markov
processes, and shows that the fixed point may be simply super-exponential
through solving the system of nonlinear equations. Note that supermarket
models are a class of complicated queueing systems and their analysis can not
apply popular queueing theory, it is necessary in the study of supermarket
models to summarize such a more general framework which enables us to focus on
important research issues. On this line, this paper develops matrix-analytical
methods of Markovian supermarket models. We hope this will be able to open a
new avenue in performance evaluation of supermarket models by means of
matrix-analytical methods.

\vskip             0.5cm


\end{abstract}

\section{Introduction}

In the study of Markovian supermarket models, this paper proposes a more
general framework including a system of differential equations for the
fraction measure and a system of nonlinear equations for the fixed point, and
the both systems of equations enable us to focus on important research issues
of Markovian supermarket models. At the same time, this paper indicates that
it is difficult and challenging to analyze the system of differential
equations and to solve the system of nonlinear equations from four key
directions: Existence of solution, uniqueness of solution, stability of
solution and effective algorithms. Since there is a large gap to provide a
complete solution to the both systems of equations, this paper devotes
heuristic understanding of how to organize and solve the both systems of
equations by means of discussing three important supermarket examples: M/G/1
type, GI/M/1 type and multiple choices. Specifically, the supermarket examples
show a key result that the fixed point can be super-exponential for more
supermarket models. Note that supermarket models are a class of complicated
queueing systems and their analysis can not apply popular queueing theory,
while recent research gave some simple and beautiful results for special
supermarket models, e.g., see Mitzenmacher \cite{Mit:1996}, Li and Lui
\cite{LiL:2010} and Luczak and McDiarmid \cite{Luc:2006}, this motivates us in
this paper to summarize a more general framework in order to develop
matrix-analytical methods of Markovian supermarket models. We hope this is
able to open a new avenue for performance evaluation of supermarket models by
means of matrix-analytical methods.

Recently, a number of companies, such as Amazon and Google, are offering cloud
computing service and cloud manufacturing technology. This motivates us in
this paper to study randomized load balancing for large-scale networks with
many computational and manufacturing resources. Randomized load balancing,
where a job is assigned to a server from a small subset of randomly chosen
servers, is very simple to implement. It can surprisingly deliver better
performance (for example reducing collisions, waiting times and backlogs) in a
number of applications including data centers, distributed memory machines,
path selection in computer networks, and task assignment at web servers.
Supermarket models are extensively used to study randomized load balancing
schemes. In the past ten years, supermarket models have been studied by
queueing theory as well as Markov processes. Since the study of supermarket
models can not apply popular queueing theory, they have not been extensively
studied in queueing committee up to now. Therefore, this leads to that
available queueing results of supermarket models are few up to now. Some
recent works dealt with the supermarket model with Poisson arrivals and
exponential service times by means of density-dependent jump Markov processes,
discussed limiting behavior of the supermarket model under a weakly convergent
setting when the population size goes to infinite, and indicated that there
exists a doubly exponential solution to the fixed point through solving the
system of nonlinear equations. Readers may refer to population processes by
Kurtz \cite{Kur:1981}, and doubly exponential solution with exponential
improvement by Vvedenskaya, Dobrushin and Karpelevich \cite{Vve:1996},
Mitzenmacher \cite{Mit:1996}, Li and Lui \cite{LiL:2010} and Luczak and
McDiarmid \cite{Luc:2006}.

Certain generalization of supermarket models has been explored in, for
example, studying simple variations by Vvedenskaya and Suhov \cite{Vve:1997},
Mitzenmacher \cite{Mit:1999}, Azar, Broder, Karlin and Upfal \cite{Azar:1999},
V\"{o}cking \cite{Voc:1999}, Mitzenmacher, Richa, and Sitaraman
\cite{Mit:2001} and Li, Lui and Wang \cite{LiLW:2010b}; considering
non-Poisson arrivals or non-exponential service times by Li, Lui and Wang
\cite{LiLW:2010a}, Li and Lui \cite{LiL:2010}, Bramson, Lu and Prabhakar
\cite{Bra:2010} and Li \cite{LiG:2010}; discussing load information by
Mirchandaney, Towsley, and Stankovic \cite{Mir:1989}, Dahlin \cite{Dah:1999}
and Mitzenmacher \cite{Mit:2000}; mathematical analysis by Graham
\cite{Gra:2000a, Gra:2000b, Gra:2004}, Luczak and Norris \cite{LucN:2005} and
Luczak and McDiarmid \cite{Luc:2006, Luc:2007}; using fast Jackson networks by
Martin and Suhov \cite{Mar:1999}, Martin \cite{Mar:2001} and Suhov and
Vvedenskaya \cite{Suh:2002}.

The main contributions of the paper are twofold. The first one is to propose a
more general framework for Marovian supermarket models. This framework
contains a system of differential equations for the fraction measure and a
system of nonlinear equations for the fixed point. It is indicated that there
exist more difficulties and challenges for dealing with the system of
differential equations and for solving the system of nonlinear equations
because of two key factors: infinite dimension and complicated structure of
nonlinear equations. Since there is still a large gap up to being able to deal
with the both systems of equations systematically, the second contribution of
this paper is to analyze three important supermarket examples: M/G/1 type,
GI/M/1 type and multiple choices. These examples provide necessary
understanding and heuristic methods in order to discuss the both systems of
equations from practical and more general point of view. For the supermarket
examples, this paper derives the systems of differential equations for the
fraction measure by means of density-dependent jump Markov processes, and
illustrates that the fixed points may be super-exponential through solving the
systems of nonlinear equations by means of matrix-analytic methods.

The remainder of this paper is organized as follows. Section 2 proposes a more
general framework for Markovian supermarket models. This framework contains a
system of differential equations for the fraction measure and a system of
nonlinear equations for the fixed point. In Sections 3 and 4, we consider a
supermarket model of M/G/1 type by means of BMAPs and a supermarket model of
GI/M/1 type in terms of batch PH service processes, respectively. For the both
supermarket models, we derive the systems of differential equations satisfied
by the fraction measure in terms of density-dependent jump Markov processes,
and obtain the system of nonlinear equations satisfied by the fixed point
which is shown to be super-exponential. In Section 5, we analyze two
supermarket models with multiple choice numbers, and give super-exponential
solution to the fixed points for the two supermarket models. Note that the
supermarket examples discussed in Sections 3 to 5 can provide a heuristic
understanding for the more general framework of Markovian supermarket model
given in Section 2.

\section{Markovian Supermarket Models}

In this section, we propose a more general framework for Markovian supermarket
models. This framework contains a system of differential equations for the
fraction measure and a system of nonlinear equations for the fixed point.

Recent research, e.g., see Mitzenmacher \cite{Mit:1999} and Li and Lui
\cite{LiL:2010}, shows that a Markovian supermarket model contains two
important factors:

(1) Continuous-time Markov chain $Q$, called stochastic environment of the
supermarket model; and

(2) Choice numbers, including input choice numbers $d_{1},d_{2},\ldots,d_{v}$
and output choice numbers $f_{1},f_{2},\ldots,f_{w}$. Note that the choice
numbers determine decomposed structure of the stochastic environment $Q$.

We first analyze stochastic environment of the Markovian supermarket model.
From point of view of stochastic models, we take a more general stochastic
environment which is a continuous-time Markov chain $\left\{  X_{t}%
,t\geq0\right\}  $ with block structure. We assume that the Markov chain
$\left\{  X_{t},t\geq0\right\}  $ on state space $\Omega=\left\{  \left(
k,j\right)  :k\geq0,1\leq j\leq m_{k}\right\}  $ is irreducible and positive
recurrent, and that its infinitesimal generator is given by%
\begin{equation}
Q=\left(
\begin{array}
[c]{cccc}%
Q_{0,0} & Q_{0,1} & Q_{0,2} & \cdots\\
Q_{1,0} & Q_{1,1} & Q_{1,2} & \cdots\\
Q_{2,0} & Q_{2,1} & Q_{2,2} & \cdots\\
\vdots & \vdots & \vdots & \ddots
\end{array}
\right)  , \label{EqG1}%
\end{equation}
where $Q_{i,j}$ is a matrix of size $m_{i}\times m_{j}$ whose $\left(
r,r^{\prime}\right)  $th entry is the transition rate of the Markov chain from
state $\left(  i,r\right)  $ to state $\left(  j,r^{\prime}\right)  $. It is
well-known that $Q_{i,j}\geq0$ for $i\neq j$, $Q_{i,i}$ is invertible with
strictly negative diagonal entries and nonnegative off-diagonal entries. For
state $\left(  i,k\right)  $, $i$ is called the \textit{level variable} and
$k$ the \textit{phase variable}. We write level $i$ as $L_{i}=\left\{  \left(
i,k\right)  :1\leq k\leq m_{i}\right\}  $.

Since the Markov chain is irreducible, for each level $i$ there must exist at
east one left-block state transition: $\leftarrow$ level $i$ or level
$i\leftarrow$, and at east one right-block state transition: $\rightarrow$
level $i$ or level $i\rightarrow$. We write%
\[
E_{\text{left}}=\left\{  \leftarrow\text{level }i\text{ or level }%
i\leftarrow:\text{level }i\in\Omega\right\}
\]
and%
\[
E_{\text{right}}=\left\{  \rightarrow\text{level }i\text{ or level
}i\rightarrow:\text{level }i\in\Omega\right\}  .
\]
Note that $E_{\text{left}}$ and $E_{\text{right}}$ describe output and input
processes in the supermarket model. Based on the two block-transition sets
$E_{\text{left}}$ and $E_{\text{right}}$, we write%
\begin{equation}
Q=Q_{\text{left}}+Q_{\text{right}}, \label{EqG2}%
\end{equation}
and for $i\geq0$%
\[
Q_{i,i}=Q_{i}^{\text{left}}+Q_{i}^{\text{right}}.
\]
Thus we have%
\[
Q_{\text{left}}=\left(
\begin{array}
[c]{cccc}%
Q_{0}^{\text{left}} &  &  & \\
Q_{1,0} & Q_{1}^{\text{left}} &  & \\
Q_{2,0} & Q_{2,1} & Q_{2}^{\text{left}} & \\
\vdots & \vdots & \vdots & \ddots
\end{array}
\right)
\]
and%
\[
Q_{\text{right}}=\left(
\begin{array}
[c]{cccc}%
Q_{0}^{\text{right}} & Q_{0,1} & Q_{0,2} & \cdots\\
& Q_{1}^{\text{right}} & Q_{1,2} & \cdots\\
&  & Q_{2}^{\text{right}} & \cdots\\
&  &  & \ddots
\end{array}
\right)  .
\]
Note that $Qe=0,Q_{\text{left}}e=0$ and $Q_{\text{right}}e=0$, where $e$ is a
column vector of ones with a suitable dimension in the context. We assume that
the matrices $Q_{j}^{\text{left}}$ for $j\geq1$ and $Q_{i}^{\text{right}}$ for
$i\geq0$ are all invertible, while $Q_{0}^{\text{left}}$ is possibly singular
if there is not an output process in level $0$. We call $Q=Q_{\text{left}%
}+Q_{\text{right}}$ an input-output rate decomposition of the Markovian
supermarket model.

Now, we provide a choice decomposition of the Markovian supermarket model
through decomposing the two matrices $Q_{\text{left}}$ and $Q_{\text{right}}$.
Note that the choice decomposition is based on the input choice numbers
$d_{1},d_{2},\ldots,d_{v}$ and the output choice numbers $f_{1},f_{2}%
,\ldots,f_{w}$. We write%
\begin{equation}
Q_{\text{left}}=Q_{\text{left}}\left(  f_{1}\right)  +Q_{\text{left}}\left(
f_{2}\right)  +\cdots+Q_{\text{left}}\left(  f_{w}\right)  \label{EqG4}%
\end{equation}
for the output choice numbers $f_{1},f_{2},\ldots,f_{w}$, and%
\begin{equation}
Q_{\text{right}}=Q_{\text{right}}\left(  d_{1}\right)  +Q_{\text{right}%
}\left(  d_{2}\right)  +\cdots+Q_{\text{right}}\left(  d_{v}\right)
\label{EqG3}%
\end{equation}
for the input choice numbers $d_{1},d_{2},\ldots,d_{v}$.

To study the Markovian supermarket model, we need to introduce two vector
notation. For a vector $a=\left(  a_{1},a_{2},a_{3},\ldots\right)  $, we write%
\[
a^{\odot d}=\left(  a_{1}^{d},a_{2}^{d},a_{3}^{d},\ldots\right)
\]
and%
\[
a^{\odot\frac{1}{d}}=\left(  a_{1}^{\frac{1}{d}},a_{2}^{\frac{1}{d}}%
,a_{3}^{\frac{1}{d}},\ldots\right)  .
\]

Let $S\left(  t\right)  =\left(  S_{0}\left(  t\right)  ,S_{1}\left(
t\right)  ,S_{2}\left(  t\right)  ,\ldots\right)  $ be the fraction measure of
the Markovian supermarket model, where $S_{i}\left(  t\right)  $ is a row
vector of size $m_{i}$ for $i\geq0$. Then $S\left(  t\right)  \geq0$ and
$S_{0}\left(  t\right)  e=1$. Based on the input-output rate decomposition and
the choice decomposition for the stochastic environment, we introduce the
following system of differential equations satisfied by the fraction measure
$S\left(  t\right)  $ as follows:%
\begin{equation}
S_{0}\left(  t\right)  \geq0\text{ and }S_{0}\left(  t\right)  e=1,
\label{EqG5}%
\end{equation}
and%
\begin{equation}
\frac{\text{d}}{\text{d}t}S\left(  t\right)  =\sum_{l=1}^{w}S^{\odot f_{l}%
}\left(  t\right)  Q_{\text{left}}\left(  f_{l}\right)  +\sum_{k=1}%
^{v}S^{\odot d_{k}}\left(  t\right)  Q_{\text{right}}\left(  d_{k}\right)  .
\label{EqG6}%
\end{equation}

In the Markovian supermarket model, a row vector $\pi=\left(  \pi_{0},\pi
_{1},\pi_{2},\ldots\right)  $ is called a fixed point of the fraction measure
$S\left(  t\right)  $ if $\lim_{t\rightarrow+\infty}S\left(  t\right)  =\pi$.
In this case, it is easy to see that%
\[
\lim_{t\rightarrow+\infty}\left[  \frac{\mathtt{d}}{\text{d}t}S\left(
t\right)  \right]  =0.
\]

If there exists a fixed point of the fraction measure, then it follows from
(\ref{EqG5}) and (\ref{EqG6}) that the fixed point is a nonnegative non-zero
solution to the following system of nonlinear equations%
\begin{equation}
\pi_{0}\geq0\text{ and }\pi_{0}e=1, \label{EqG7}%
\end{equation}
and%
\begin{equation}
\sum_{l=1}^{w}\pi^{\odot f_{l}}Q_{\text{left}}\left(  f_{l}\right)
+\sum_{k=1}^{v}\pi^{\odot d_{k}}Q_{\text{right}}\left(  d_{k}\right)  =0.
\label{EqG8}%
\end{equation}

\begin{Rem}
If $d_{k}=1$ for $1\leq k\leq v$ and $f_{l}=1$ for $1\leq l\leq w$, then the
system of differential equations (\ref{EqG5}) and (\ref{EqG6}) is given by%
\[
S_{0}\left(  t\right)  \geq0\text{ and }S_{0}\left(  t\right)  e=1,
\]
and%
\[
\frac{\text{d}}{\text{d}t}S\left(  t\right)  =S\left(  t\right)  Q.
\]
Thus we obtain%
\[
S\left(  t\right)  =cS\left(  0\right)  \exp\left\{  Qt\right\}  .
\]
Let%
\[
W\left(  t\right)  =\left(  W_{0}\left(  t\right)  ,W_{1}\left(  t\right)
,W_{2}\left(  t\right)  ,\ldots\right)  =S\left(  0\right)  \exp\left\{
Qt\right\}  ,
\]
where $W_{i}\left(  t\right)  $ is a row vector of size $m_{i}$ for $i\geq0$.
Then $S\left(  t\right)  =cW\left(  t\right)  ,$ where $c=1/W_{0}\left(
t\right)  e$. At the same time, the system of nonlinear equations (\ref{EqG7})
and (\ref{EqG8}) is given by%
\[
\pi_{0}\geq0\text{ and }\pi_{0}e=1,
\]
and%
\[
\pi Q=0.
\]
Let $W=\left(  w_{0},w_{1},w_{2},\ldots\right)  $ be the stationary
probability vector of the Markov chain $Q$, where $W_{i}$ is a row vector of
size $m_{i}$ for $i\geq0$. Then $\pi=cW$, where $c=1/w_{0}e$. Note that the
stationary probability vector $W$ of the block-structured Markov chain $Q$\ is
given a detailed analysis in Chapter 2 of Li \cite{Li:2010} by means of the RG-factorizations.
\end{Rem}

If there exist some $d_{k}\geq2$ or/and $f_{l}\geq2$ in the Markovian
supermarket model, then the system of differential equations (\ref{EqG5}) and
(\ref{EqG6}) and the system of nonlinear equations (\ref{EqG7}) to
(\ref{EqG8}) are two decomposed power-form generalizations of transient
solution and of stationary probability of an irreducible continuous-time
Markov chain with block structure (see Chapters 2 and 8 of Li \cite{Li:2010}).
Note that Li \cite{Li:2010} can deal with transient solution and stationary
probability for an irreducible block-structured Markov chain, where the
RG-factorizatons play a key role. However, the RG-factorizatons can not hold
for Markovian supermarket models with some $d_{k}\geq2$ or/and $f_{l}\geq2$.
Therefore, there exist more difficulties and challenges to study the system of
differential equations (\ref{EqG5}) and (\ref{EqG6}) and the system of
nonlinear equations (\ref{EqG7}) to (\ref{EqG8}). Specifically, it still keeps
not to be able to answer four important issues: Existence of solution,
uniqueness of solution, stability of solution and effective algorithms. This
is similar to some research on the four important issues of irreducible
continuous-time Markov chains with block structure.

In the remainder of this paper, we will study thee important Markovian
supermarket examples: M/G/1 type, GI/M/1 type and multiple choices. Our
purpose is to provide heuristic understanding of how to set up and solve the
system of differential equations (\ref{EqG5}) and (\ref{EqG6}), and the system
of nonlinear equations (\ref{EqG7}) to (\ref{EqG8}).

\section{A Supermarket Model of M/G/1 Type}

In this section, we consider a supermarket model with a BMAP and exponential
service times. Note that the stochastic environment is a Markov chain of M/G/1
type, the supermarket model is called to be of M/G/1 type. For the supermarket
model of M/G/1 type, we set up the system of differential equations for the
fraction measure by means of density-dependent jump Markov processes, and
derive the system of nonlinear equations satisfied by the fixed point which is
shown to be super-exponential solution.

The supermarket model of M/G/1 type is described as follows. Customers arrive
at a queueing system of $n>1$ servers as a BMAP with irreducible matrix
descriptor $\left(  nC,nD_{1},nD_{2},nD_{3},\ldots\right)  $ of size $m$,
where the matrix $C$ is invertible and has strictly negative diagonal entries
and nonnegative off-diagonal; $D_{k}\geq0$ is the arrival rate matrix with
batch size $k$ for $k\geq1$. We assume that $\sum_{k=1}^{\infty}kD_{k}$ is
finite and that $C+\sum_{k=1}^{\infty}D_{k}$ is an irreducible infinitesimal
generator with $\left(  C+\sum_{k=1}^{\infty}D_{k}\right)  e=0$. Let $\gamma$
be the stationary probability vector of the irreducible Markov chain
$C+\sum_{k=1}^{\infty}D_{k}$. Then the stationary arrival rate of the BMAP is
given by $n\lambda=n\gamma\sum_{k=1}^{\infty}kD_{k}e$. The service times of
each customer are exponentially distributed with service rate $\mu$. Each
batch of arriving customers choose $d\geq1$ servers independently and
uniformly at random from the $n$ servers, and joins for service at the server
which currently possesses the fewest number of customers. If there is a tie,
servers with the fewest number of customers will be chosen randomly. All
customers in every server will be served in the first-come-first service
(FCFS) manner. We assume that all the random variables defined above are
independent of each other and that this system is operating in the region
$\rho=\lambda/\mu<1$. Clearly, $d$ is an input choice number in this
supermarket model. Figure 1 is depicted as an illustration for supermarket
models of M/G/1 type.

\begin{figure}[ptbh]
\centering                         \includegraphics[width=8cm]{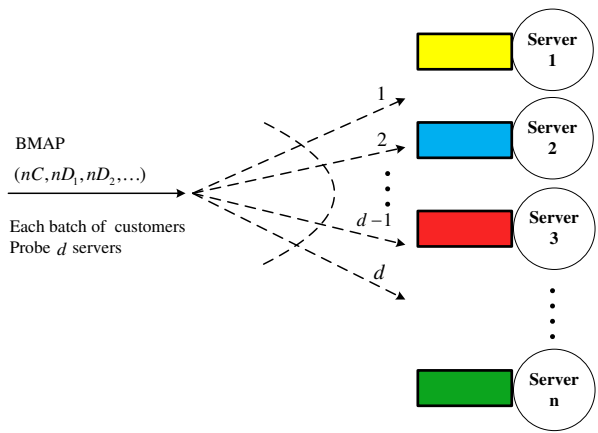}
\caption{A supermarket model of M/G/1 type}%
\label{figure: model-1}%
\end{figure}

The supermarket model with a BMAP and exponential service times is stable if
$\rho=\lambda/\mu<1$. This proof can be given by a simple comparison argument
with the queueing system in which each customer queues at a random server
(i.e., where $d=1$). When $d=1$, each server acts like a BMAP/M/1 queue which
is stable if $\rho=\lambda/\mu<1$, see chapter 5 in Neuts \cite{Neu:1989}.
Similar to analysis in Winston \cite{Win:1977} and Weber \cite{Web:1978}, the
comparison argument leads to two useful results: (1) the shortest queue is
optimal due to the assumptions on a BMAP and exponential service times in the
supermarket model; and (2) the size of the longest queue in the supermarket
model is stochastically dominated by the size of the longest queue in a set of
$n$ independent BMAP/M/1 queues.

We define $n_{k}^{\left(  i\right)  }\left(  t\right)  $ as the number of
queues with at least $k$ customers, including customers in service, and with
the BMAP in phase $i$\ at time $t\geq0$. Clearly, $0\leq n_{k}^{\left(
i\right)  }\left(  t\right)  \leq n$ for $k\geq0$ and $1\leq i\leq m$. Let%
\[
x_{n}^{\left(  i\right)  }\left(  k,t\right)  =\frac{n_{k}^{\left(  i\right)
}\left(  t\right)  }{n},
\]
which is the fraction of queues with at least $k$ customers and the BMAP in
phase $i$\ at time $t\geq0$ for $k\geq0$. We write%
\[
X_{n}\left(  k,t\right)  =\left(  x_{n}^{\left(  1\right)  }\left(
k,t\right)  ,x_{n}^{\left(  2\right)  }\left(  k,t\right)  ,\ldots
,x_{n}^{\left(  m\right)  }\left(  k,t\right)  \right)
\]
for $k\geq0$, and%
\[
X_{n}\left(  t\right)  =\left(  X_{n}\left(  0,t\right)  ,X_{n}\left(
1,t\right)  ,X_{n}\left(  2,t\right)  ,\ldots\right)  .
\]
The state of the supermarket model may be described by the vector
$X_{n}\left(  t\right)  $ for $t\geq0$. Since the arrival process to the
queueing system is a BMAP and the service time of each customer is
exponential, the stochastic process $\left\{  X_{n}\left(  t\right)
,t\geq0\right\}  $ is a Markov process whose state space is given by%
\begin{align*}
\Omega_{n}=  &  \{\left(  g_{n}^{\left(  0\right)  },g_{n}^{\left(  1\right)
},g_{n}^{\left(  2\right)  }\ldots\right)  :g_{n}^{\left(  0\right)  }\text{
is a probability vector, }g_{n}^{\left(  k\right)  }\geq g_{n}^{\left(
k+1\right)  }\geq0\\
&  \text{ for }k\geq1\text{,}\text{ and }ng_{n}^{\left(  l\right)  }\text{ is
a vector of nonnegative integers for }l\geq0\}.
\end{align*}
Let%
\[
s_{k}^{\left(  i\right)  }\left(  n,t\right)  =E\left[  x_{k}^{\left(
i\right)  }\left(  n,t\right)  \right]  ,
\]
and%
\[
S_{k}\left(  n,t\right)  =\left(  s_{k}^{\left(  1\right)  }\left(
n,t\right)  ,s_{k}^{\left(  2\right)  }\left(  n,t\right)  ,\ldots
,s_{k}^{\left(  m\right)  }\left(  n,t\right)  \right)
\]
for $k\geq0,$%
\[
S\left(  n,t\right)  =\left(  S_{0}\left(  n,t\right)  ,S_{1}\left(
n,t\right)  ,S_{2}\left(  n,t\right)  ,\ldots\right)  .
\]
As shown in Martin and Suhov \cite{Mar:1999} and Luczak and McDiarmid
\cite{Luc:2006}, the Markov process $\left\{  X_{n}\left(  t\right)
,t\geq0\right\}  $ is asymptotically deterministic as $n\rightarrow\infty$.
Thus $\lim_{n\rightarrow\infty}E\left[  x_{k}^{\left(  i\right)  }\left(
n,t\right)  \right]  $ always exist by means of the law of large numbers for
$k\geq0$. Based on this, we write%
\[
S_{k}\left(  t\right)  =\lim_{n\rightarrow\infty}S_{k}\left(  n,t\right)
\]
for $k\geq0$, and%
\[
S\left(  t\right)  =\left(  S_{0}\left(  t\right)  ,S_{1}\left(  t\right)
,S_{2}\left(  t\right)  ,\ldots\right)  .
\]
Let $X\left(  t\right)  =\lim_{n\rightarrow\infty}X_{n}\left(  t\right)  $.
Then it is easy to see from the BMAP and the exponential service times that
$\left\{  X\left(  t\right)  ,t\geq0\right\}  $ is also a Markov process whose
state space is given by%
\[
\Omega=\left\{  \left(  g^{\left(  0\right)  },g^{\left(  1\right)
},g^{\left(  2\right)  },\ldots\right)  :g^{\left(  0\right)  }\text{ is a
probability vector},g^{\left(  k\right)  }\geq g^{\left(  k+1\right)  }%
\geq0\text{ for }k\geq1\right\}  .
\]
If the initial distribution of the Markov process $\left\{  X_{n}\left(
t\right)  ,t\geq0\right\}  $ approaches the Dirac delta-measure concentrated
at a point $g\in$ $\Omega$, then $X\left(  t\right)  =\lim_{n\rightarrow
\infty}X_{n}\left(  t\right)  $ is concentrated on the trajectory
$S_{g}=\left\{  S\left(  t\right)  :t\geq0\right\}  $. This indicates a law of
large numbers for the time evolution of the fraction of queues of different
lengths. Furthermore, the Markov process $\left\{  X_{n}\left(  t\right)
,t\geq0\right\}  $ converges weakly to the fraction vector $S\left(  t\right)
=\left(  S_{0}\left(  t\right)  ,S_{1}\left(  t\right)  ,S_{2}\left(
t\right)  ,\ldots\right)  $ as $n\rightarrow\infty$, or for a sufficiently
small $\varepsilon>0$,%
\[
\lim_{n\rightarrow\infty}P\left\{  ||X_{n}\left(  t\right)  -S\left(
t\right)  ||\geq\varepsilon\right\}  =0,
\]
where $||a||$ is the $L_{\infty}$-norm of vector $a$.

In what follows we set up a system of differential vector equations satisfied
by the fraction vector $S\left(  t\right)  $ by means of density-dependent
jump Markov processes.

We first provide an example to indicate how to derive the differential vector
equations. Consider the supermarket model with $n$ servers, and determine the
expected change in the number of queues with at least $k$ customers over a
small time interval $[0,dt)$. The probability vector that an arriving customer
joins a queue with $k-1$ customers in this time interval is given by%
\[
\left[  S_{0}^{\odot d}\left(  n,t\right)  D_{k}+S_{1}^{\odot d}\left(
n,t\right)  D_{k-1}+\cdots+S_{k-1}^{\odot d}\left(  n,t\right)  D_{1}%
+S_{k}^{\odot d}\left(  n,t\right)  C\right]  \cdot n\text{d}t,
\]
since each arriving customer chooses $d$ servers independently and uniformly
at random from the $n$ servers, and waits for service at the server which
currently contains the fewest number of customers. Similarly, the probability
vector that a customer leaves a server queued by $k$ customers in this time
interval is given by%
\[
\left[  -\mu S_{k}\left(  n,t\right)  +\mu S_{k+1}\left(  n,t\right)  \right]
\cdot n\text{d}t.
\]
Therefore, we obtain%
\begin{align*}
\text{d}E\left[  n_{k}\left(  n,t\right)  \right]  =  &  \left[  \sum
_{l=0}^{k-1}S_{l}^{\odot d}\left(  n,t\right)  D_{k-l}+S_{k}^{\odot d}\left(
n,t\right)  C\right]  \cdot n\text{d}t\\
&  +\left[  -\mu S_{k}\left(  n,t\right)  +\mu S_{k+1}\left(  n,t\right)
\right]  \cdot n\text{d}t.
\end{align*}
This leads to%
\begin{equation}
\frac{\text{d}S_{k}\left(  n,t\right)  }{\text{d}t}=\sum_{l=0}^{k-1}%
S_{l}^{\odot d}\left(  n,t\right)  D_{k-l}+S_{k}^{\odot d}\left(  n,t\right)
C+-\mu S_{k}\left(  n,t\right)  +\mu S_{k+1}\left(  n,t\right)  . \label{Eq-2}%
\end{equation}
Since $\lim_{n\rightarrow\infty}E\left[  x_{k}^{\left(  i\right)  }\left(
n,t\right)  \right]  $ always exists for $k\geq0$, taking $n\rightarrow\infty$
in both sides of Equation (\ref{Eq-2}) we can easily obtain%
\begin{equation}
\frac{\text{d}S_{k}\left(  t\right)  }{\text{d}t}=\sum_{l=0}^{k-1}S_{l}^{\odot
d}\left(  t\right)  D_{k-l}+S_{k}^{\odot d}\left(  t\right)  C-\mu
S_{k}\left(  t\right)  +\mu S_{k+1}\left(  t\right)  . \label{Eq-1}%
\end{equation}

Using a similar analysis to that in Equation (\ref{Eq-1}), we obtain the
system of differential vector equations for the fraction vector $S\left(
t\right)  =\left(  S_{0}\left(  t\right)  ,S_{1}\left(  t\right)
,\ldots\right)  $ as follows:%
\begin{equation}
S_{0}\left(  t\right)  \geq0,S_{0}\left(  t\right)  e=1, \label{Eq0}%
\end{equation}%
\begin{equation}
\frac{\mathtt{d}}{\text{d}t}S_{0}\left(  t\right)  =S_{0}^{\odot d}\left(
t\right)  C+\mu S_{1}\left(  t\right)  \label{Eq1}%
\end{equation}
and for $k\geq1$%
\begin{equation}
\frac{\mathtt{d}}{\text{d}t}S_{k}\left(  t\right)  =\sum_{l=0}^{k-1}%
S_{l}^{\odot d}\left(  n,t\right)  D_{k-l}+S_{k}^{\odot d}\left(  n,t\right)
C-\mu S_{k}\left(  n,t\right)  +\mu S_{k+1}\left(  n,t\right)  . \label{Eq2}%
\end{equation}

Let $\pi$ be the fixed point. Then $\pi$ satisfies the following system of
nonlinear equations%
\begin{equation}
\pi_{0}\geq0,\pi_{0}e=1, \label{EqE2-1}%
\end{equation}%
\begin{equation}
\pi_{0}^{\odot d}C+\mu\pi_{1}=0 \label{EqE2-2}%
\end{equation}
and for $k\geq1$,%
\begin{equation}
\sum_{l=0}^{k-1}\pi_{l}^{\odot d}D_{k-l}+\pi_{k}^{\odot d}C-\mu\pi_{k}+\mu
\pi_{k+1}=0. \label{EqE2-3}%
\end{equation}

Let%
\[
Q_{\text{right}}=\left(
\begin{array}
[c]{cccccc}%
C & D_{1} & D_{2} & D_{3} & D_{4} & \cdots\\
& C & D_{1} & D_{2} & D_{3} & \cdots\\
&  & C & D_{1} & D_{2} & \cdots\\
&  &  & C & D_{1} & \cdots\\
&  &  &  & \ddots &
\end{array}
\right)
\]
and%
\[
Q_{\text{left}}=\left(
\begin{array}
[c]{ccccc}%
0 &  &  &  & \\
\mu I & -\mu I &  &  & \\
& \mu I & -\mu I &  & \\
&  & \mu I & -\mu I & \\
&  &  & \ddots & \ddots
\end{array}
\right)  .
\]
Then the system of differential vector equations is given by%
\[
S_{0}\left(  t\right)  \geq0,S_{0}\left(  t\right)  e=1,
\]
and%
\[
\frac{\text{d}}{\text{d}t}S\left(  t\right)  =S^{\odot d}\left(  t\right)
Q_{\text{right}}+S\left(  t\right)  Q_{\text{left}};
\]
and the system of nonlinear equations (\ref{EqE2-1}) to (\ref{EqE2-3}) is
given by%
\[
\pi_{0}\geq0,\pi_{0}e=1,
\]
and%
\[
\pi^{\odot d}Q_{\text{right}}+\pi Q_{\text{left}}=0.
\]

\begin{Rem}
For the supermarket model with a BMAP and exponential service times, its
stochastic environment is a Markov chain of M/G/1 type whose infinitesimal
generator is given by $Q=Q_{\text{left}}+Q_{\text{right}}$. This example
clearly indicates how to set up the system of differential equations
(\ref{EqG5}) and (\ref{EqG6}) for the fraction measure and the system of
nonlinear equations (\ref{EqG7}) to (\ref{EqG8}) for the fixed point.
\end{Rem}

In the remainder of this section, we provide a super-exponential solution to
the fixed point $\pi$ by means of some useful relations among the vectors
$\pi_{k}$ for $k\geq0$.

It follows from (\ref{EqE2-3}) that
\begin{align}
&  \left(  \pi_{1}^{\odot d},\pi_{2}^{\odot d},\pi_{3}^{\odot d}%
,\ldots\right)  \left(
\begin{array}
[c]{cccc}%
C & D_{1} & D_{2} & \cdots\\
& C & D_{1} & \cdots\\
&  & C & \cdots\\
&  &  & \ddots
\end{array}
\right)  +\left(  \pi_{1},\pi_{2},\pi_{3},\ldots\right)  \left(
\begin{array}
[c]{cccc}%
-\mu I &  &  & \\
\mu I & -\mu I &  & \\
& \mu I & -\mu I & \\
&  & \ddots & \ddots
\end{array}
\right) \nonumber\\
&  =-\left(  \pi_{0}^{\odot d}D_{1},\pi_{0}^{\odot d}D_{2},\pi_{0}^{\odot
d}D_{3},\ldots\right)  . \label{EqE2-7}%
\end{align}
Let%
\[
A=\left(
\begin{array}
[c]{cccc}%
-\mu I &  &  & \\
\mu I & -\mu I &  & \\
& \mu I & -\mu I & \\
&  & \ddots & \ddots
\end{array}
\right)  .
\]
Then%
\[
\left(  -A\right)  ^{-1}=\left(
\begin{array}
[c]{cccc}%
\frac{1}{\mu}I &  &  & \\
\frac{1}{\mu}I & \frac{1}{\mu}I &  & \\
\frac{1}{\mu}I & \frac{1}{\mu}I & \frac{1}{\mu}I & \\
\vdots & \vdots & \vdots & \ddots
\end{array}
\right)  .
\]
Note that%
\[
\left(
\begin{array}
[c]{cccc}%
D_{0} & D_{1} & D_{2} & \cdots\\
& D_{0} & D_{1} & \cdots\\
&  & D_{0} & \cdots\\
&  &  & \ddots
\end{array}
\right)  \left(  -A^{-1}\right)  =\left(
\begin{array}
[c]{cccc}%
\frac{1}{\mu}\sum\limits_{k=0}^{\infty}D_{k} & \frac{1}{\mu}\sum
\limits_{k=1}^{\infty}D_{k} & \frac{1}{\mu}\sum\limits_{k=2}^{\infty}D_{k} &
\cdots\\
\frac{1}{\mu}\sum\limits_{k=0}^{\infty}D_{k} & \frac{1}{\mu}\sum
\limits_{k=0}^{\infty}D_{k} & \frac{1}{\mu}\sum\limits_{k=1}^{\infty}D_{k} &
\cdots\\
\frac{1}{\mu}\sum\limits_{k=0}^{\infty}D_{k} & \frac{1}{\mu}\sum
\limits_{k=0}^{\infty}D_{k} & \frac{1}{\mu}\sum\limits_{k=0}^{\infty}D_{k} &
\cdots\\
\vdots & \vdots & \vdots &
\end{array}
\right)
\]
and%
\[
\left(  \pi_{0}^{\odot d}D_{1},\pi_{0}^{\odot d}D_{2},\pi_{0}^{\odot d}%
D_{3},\ldots\right)  \left(  -A^{-1}\right)  =\left(  \frac{1}{\mu}\pi
_{0}^{\odot d}\sum\limits_{k=1}^{\infty}D_{k},\frac{1}{\mu}\pi_{0}^{\odot
d}\sum\limits_{k=2}^{\infty}D_{k},\frac{1}{\mu}\pi_{0}^{\odot d}%
\sum\limits_{k=3}^{\infty}D_{k},\ldots\right)  ,
\]
it follows from (\ref{EqE2-7}) that%
\begin{equation}
\pi_{1}=\pi_{0}^{\odot d}\left[  \frac{1}{\mu}\sum\limits_{i=1}^{\infty}%
D_{i}\right]  +\sum_{j=1}^{\infty}\pi_{j}^{\odot d}\left[  \frac{1}{\mu}%
\sum\limits_{i=0}^{\infty}D_{i}\right]  \label{EqE2-8}%
\end{equation}
and for $k\geq2$,%
\begin{equation}
\pi_{k}=\sum\limits_{i=0}^{k-1}\pi_{i}^{\odot d}\left[  \frac{1}{\mu}%
\sum\limits_{j=k-i}^{\infty}D_{j}\right]  +\sum_{j=k}^{\infty}\pi_{j}^{\odot
d}\left[  \frac{1}{\mu}\sum\limits_{i=0}^{\infty}D_{i}\right]  .
\label{EqE2-9}%
\end{equation}
To omit the terms $\sum_{j=k}^{\infty}\pi_{j}^{\odot d}\left[  \frac{1}{\mu
}\sum\limits_{i=0}^{\infty}D_{i}\right]  $ for $k\geq1$, we assume that the
system of nonlinear equations (\ref{EqE2-8}) and (\ref{EqE2-9}) has a
closed-form solution%
\begin{equation}
\pi_{k}=r\left(  k\right)  \gamma^{\odot\frac{1}{d}}, \label{EqE2-10}%
\end{equation}
where $r\left(  k\right)  $ is an underdetermined positive constant for
$k\geq1$. Then it follows from (\ref{EqE2-8}), (\ref{EqE2-9}) and
(\ref{EqE2-10}) that%
\begin{equation}
\pi_{1}=\pi_{0}^{\odot d}\left[  \frac{1}{\mu}\sum\limits_{i=1}^{\infty}%
D_{i}\right]  \label{EqE2-11-0}%
\end{equation}
or%
\begin{equation}
r\left(  1\right)  \gamma^{\odot\frac{1}{d}}=\pi_{0}^{\odot d}\left[
\frac{1}{\mu}\sum\limits_{i=1}^{\infty}D_{i}\right]  ; \label{EqE2-11}%
\end{equation}
and for $k\geq2$,%
\[
\pi_{k}=\sum\limits_{i=0}^{k-1}\pi_{i}^{\odot d}\left[  \frac{1}{\mu}%
\sum\limits_{j=k-i}^{\infty}D_{j}\right]
\]
or%
\begin{equation}
r\left(  k\right)  \gamma^{\odot\frac{1}{d}}=\pi_{0}^{\odot d}\left[
\frac{1}{\mu}\sum\limits_{j=k}^{\infty}D_{j}\right]  +\sum\limits_{i=0}%
^{k-1}\left[  r\left(  i\right)  \right]  ^{d}\gamma\left[  \frac{1}{\mu}%
\sum\limits_{j=k-i}^{\infty}D_{j}\right]  . \label{EqE2-12}%
\end{equation}
Let $\theta=1/\gamma^{\odot\frac{1}{d}}e$. Then $0<\theta<1$. Let $\lambda
_{k}=\gamma\sum_{i=k}^{\infty}D_{i}e$ and $\rho_{k}=\lambda_{k}/\mu$. Then it
follows from (\ref{EqE2-11}) and (\ref{EqE2-12}) that%
\begin{equation}
r\left(  1\right)  =\frac{\theta}{\mu}\pi_{0}^{\odot d}\sum\limits_{i=1}%
^{\infty}D_{i}e \label{EqE2-13}%
\end{equation}
and for $k\geq2$%
\begin{align}
r\left(  k\right)   &  =\frac{\theta}{\mu}\pi_{0}^{\odot d}\sum\limits_{j=k}%
^{\infty}D_{j}e+\frac{\theta}{\mu}\sum\limits_{i=1}^{k-1}\left[  r\left(
i\right)  \right]  ^{d}\gamma\sum\limits_{j=k-i}^{\infty}D_{j}e\nonumber\\
&  =\frac{\theta}{\mu}\pi_{0}^{\odot d}\sum\limits_{j=k}^{\infty}D_{j}%
e+\theta\sum\limits_{i=1}^{k-1}\left[  r\left(  i\right)  \right]  ^{d}%
\rho_{k-i}. \label{EqE2-14}%
\end{align}
It is easy to see from (\ref{EqE2-13}) and (\ref{EqE2-14}) that $\pi_{0}$ and
$r\left(  1\right)  $ are two key underdetermined terms for the closed-form
solution to the system of nonlinear equations (\ref{EqE2-11}) and
(\ref{EqE2-12}). Let us first derive the vector $\pi_{0}$. It follows from
(\ref{EqE2-2}) and (\ref{EqE2-11-0}) that%
\[
\left\{
\begin{array}
[c]{l}%
\pi_{0}^{\odot d}C+\mu\pi_{1}=0,\\
\mu\pi_{1}=\pi_{0}^{\odot d}\sum\limits_{i=1}^{\infty}D_{i}.
\end{array}
\right.
\]
This leads to%
\[
\pi_{0}^{\odot d}\left(  C+\sum\limits_{i=1}^{\infty}D_{i}\right)  =0.
\]
Thus, it is easy to see that $\pi_{0}=\theta\gamma^{\odot\frac{1}{d}}$, which
is a probability vector with $\pi_{0}e=1$. Hence we have%
\begin{equation}
\pi_{1}=-\frac{\theta^{d}}{\mu}\gamma C=\frac{\theta^{d}}{\mu}\gamma
\sum\limits_{i=1}^{\infty}D_{i}. \label{EqE2-15}%
\end{equation}
It follows from (\ref{EqE2-13}) and (\ref{EqE2-14}) that%
\begin{equation}
r\left(  1\right)  =\frac{\theta}{\mu}\cdot\theta^{d}\gamma\sum\limits_{i=1}%
^{\infty}D_{i}e=\theta^{d+1}\rho_{1} \label{EqE2-16}%
\end{equation}
and for $k\geq2$%
\begin{align}
r\left(  k\right)   &  =\frac{\theta}{\mu}\pi_{0}^{\odot d}\sum\limits_{j=k}%
^{\infty}D_{j}e+\frac{\theta}{\mu}\sum\limits_{i=1}^{k-1}\left[  r\left(
i\right)  \right]  ^{d}\gamma\sum\limits_{j=k-i}^{\infty}D_{j}e\nonumber\\
&  =\theta^{d+1}\rho_{k}+\theta\sum\limits_{i=1}^{k-1}\left[  r\left(
i\right)  \right]  ^{d}\rho_{k-i}. \label{EqE2-17}%
\end{align}
Therefore, we obtain the super-exponential solution to the fixed point as
follows:
\[
\pi_{0}=\theta\gamma^{\odot\frac{1}{d}}%
\]
and for $k\geq1$%
\[
\pi_{k}=\left[  \theta^{d+1}\rho_{k}+\theta\sum\limits_{i=1}^{k-1}\left[
r\left(  i\right)  \right]  ^{d}\rho_{k-i}\right]  \gamma^{\odot\frac{1}{d}}.
\]

\section{A Supermarket Model of GI/M/1 Type}

In this section, we analyze a supermarket model with Poisson arrivals and
batch PH service processes. Note that the stochastic environment is a Markov
chain of GI/M/1 type, thus the supermarket model is called to be of GI/M/1
type. For the supermarket model of GI/M/1 type, we set up the system of
differential equations for the fraction measure by means of density-dependent
jump Markov processes, and derive the system of nonlinear equations satisfied
the fixed point which can be computed by an iterative algorithm. Further, it
is seen that the supermarket model of GI/M/1 type is more difficult than the
case of M/G/1 type.

Let us describe the supermarket model of GI/M/1 type. Customers arrive at a
queueing system of $n>1$ servers as a Poisson process with arrival rate
$n\lambda$ for $\lambda>0$. The service times of each batch of customers are
of phase type with irreducible representation $\left(  \alpha,T\right)  $ of
order $m$ and with a batch size distribution $\left\{  b_{k},k=1,2,3,\ldots
\right\}  $ for $\sum_{k=1}^{\infty}b_{k}=1$ and $\overline{b}=\sum
_{k=1}^{\infty}kb_{k}<+\infty$. Let $T^{0}=-Te\gvertneqq0$. Then the expected
service time is given by $1/\mu=-\alpha T^{-1}e=\eta T^{0}$, where $\eta$ is
the stationary probability vector of the Markov chain $T+T^{0}\alpha$. Each
batch of arriving customers choose $d\geq1$ servers independently and
uniformly at random from the $n$ servers, and waits for service at the server
which currently contains the fewest number of customers. If there is a tie,
servers with the fewest number of customers will be chosen randomly. All
customers in every server will be served in FCFS for different batches and in
random service within one batch. We assume that all random variables defined
above are independent of each other, and that the system is operating in the
stable region $\rho=\lambda/\mu\overline{b}<1$. Clearly, $d$ is an input
choice number in this supermarket model. Figure 2 is depicted as an
illustration for supermarket models of GI/M/1 type.

\begin{figure}[ptbh]
\centering                   \includegraphics[width=8cm]{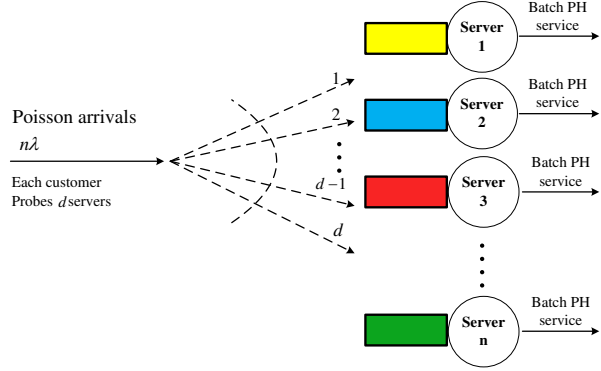}
\caption{A supermarket model of GI/M/1 type}%
\label{figure: model-2}%
\end{figure}

We define $n_{k}^{\left(  i\right)  }\left(  t\right)  $ as the number of
queues with at least $k$ customers and the service time in phase $i$ at time
$t\geq0$. Clearly, $0\leq n_{k}^{\left(  i\right)  }\left(  t\right)  \leq n$
for $k\geq1$ and $1\leq i\leq m$. Let%
\[
X_{n}^{\left(  0\right)  }\left(  t\right)  =\frac{n}{n}=1,
\]
and $k\geq1$
\[
X_{n}^{\left(  k,i\right)  }\left(  t\right)  =\frac{n_{k}^{\left(  i\right)
}\left(  t\right)  }{n},
\]
which is the fraction of queues with at least $k$ customers and the service
time in phase $i$ at time $t\geq0$. We write%
\[
X_{n}^{\left(  k\right)  }\left(  t\right)  =\left(  X_{n}^{\left(
k,1\right)  }\left(  t\right)  ,X_{n}^{\left(  k,2\right)  }\left(  t\right)
,\ldots,X_{n}^{\left(  k,m\right)  }\left(  t\right)  \right)  ,\text{
\ }k\geq1,
\]%
\[
X_{n}\left(  t\right)  =\left(  X_{n}^{\left(  0\right)  }\left(  t\right)
,X_{n}^{\left(  1\right)  }\left(  t\right)  ,X_{n}^{\left(  2\right)
}\left(  t\right)  ,\ldots\right)  .
\]
The state of the supermarket model may be described by the vector
$X_{n}\left(  t\right)  $ for $t\geq0$. Since the arrival process to the
queueing system is Poisson and the service times of each server are of phase
type, $\left\{  X_{n}\left(  t\right)  ,t\geq0\right\}  $ is a Markov process
whose state space is given by%
\begin{align*}
\Omega_{n}  &  =\{\left(  g_{n}^{\left(  0\right)  },g_{n}^{\left(  1\right)
},,g_{n}^{\left(  2\right)  }\ldots\right)  :g_{n}^{\left(  0\right)
}=1,g_{n}^{\left(  k-1\right)  }\geq g_{n}^{\left(  k\right)  }\geq0,\\
&  \text{and \ \ }ng_{n}^{\left(  k\right)  }\text{ \ is a vector of
nonnegative integers for }k\geq1\}.
\end{align*}
Let%
\[
s_{0}\left(  n,t\right)  =E\left[  X_{n}^{\left(  0\right)  }\left(  t\right)
\right]
\]
and $k\geq1$%
\[
s_{k}^{\left(  i\right)  }\left(  n,t\right)  =E\left[  X_{n}^{\left(
k,i\right)  }\left(  t\right)  \right]  .
\]
Clearly, $s_{0}\left(  n,t\right)  =1$. We write%
\[
S_{k}\left(  n,t\right)  =\left(  s_{k}^{\left(  1\right)  }\left(
n,t\right)  ,s_{k}^{\left(  2\right)  }\left(  n,t\right)  ,\ldots
,s_{k}^{\left(  m\right)  }\left(  n,t\right)  \right)  ,\text{ \ }k\geq1.
\]

As shown in Martin and Suhov \cite{Mar:1999} and Luczak and McDiarmid
\cite{Luc:2006}, the Markov process $\left\{  X_{n}\left(  t\right)
,t\geq0\right\}  $ is asymptotically deterministic as $n\rightarrow\infty$.
Thus $\lim_{n\rightarrow\infty}E\left[  X_{n}^{\left(  0\right)  }\left(
t\right)  \right]  $ and $\lim_{n\rightarrow\infty}E\left[  X_{n}^{\left(
k,i\right)  }\right]  $ always exist by means of the law of large numbers.
Based on this, we write%
\[
S_{0}\left(  t\right)  =\lim_{n\rightarrow\infty}s_{0}\left(  n,t\right)  =1,
\]
for $k\geq1$%
\[
s_{k}^{\left(  i\right)  }\left(  t\right)  =\lim_{n\rightarrow\infty}%
s_{k}^{\left(  i\right)  }\left(  n,t\right)  ,
\]%
\[
S_{k}\left(  t\right)  =\left(  s_{k}^{\left(  1\right)  }\left(  t\right)
,s_{k}^{\left(  2\right)  }\left(  t\right)  ,\ldots,s_{k}^{\left(  m\right)
}\left(  t\right)  \right)
\]
and%
\[
S\left(  t\right)  =\left(  S_{0}\left(  t\right)  ,S_{1}\left(  t\right)
,S_{2}\left(  t\right)  ,\ldots\right)  .
\]
Let $X\left(  t\right)  =\lim_{n\rightarrow\infty}X_{n}\left(  t\right)  $.
Then it is easy to see from Poisson arrivals and batch PH service times that
$\left\{  X\left(  t\right)  ,t\geq0\right\}  $ is also a Markov process whose
state space is given by%
\[
\Omega=\left\{  \left(  g^{\left(  0\right)  },g^{\left(  1\right)
},g^{\left(  2\right)  },\ldots\right)  :g^{\left(  0\right)  }=1,g^{\left(
k-1\right)  }\geq g^{\left(  k\right)  }\geq0\right\}  .
\]
If the initial distribution of the Markov process $\left\{  X_{n}\left(
t\right)  ,t\geq0\right\}  $ approaches the Dirac delta-measure concentrated
at a point $g\in$ $\Omega$, then its steady-state distribution is concentrated
in the limit on the trajectory $S_{g}=\left\{  S\left(  t\right)
:t\geq0\right\}  $. This indicates a law of large numbers for the time
evolution of the fraction of queues of different lengths. Furthermore, the
Markov process $\left\{  X_{n}\left(  t\right)  ,t\geq0\right\}  $ converges
weakly to the fraction vector $S\left(  t\right)  =\left(  S_{0}\left(
t\right)  ,S_{1}\left(  t\right)  ,S_{2}\left(  t\right)  ,\ldots\right)  $,
or for a sufficiently small $\varepsilon>0$,%
\[
\lim_{n\rightarrow\infty}P\left\{  ||X_{n}\left(  t\right)  -S\left(
t\right)  ||\geq\varepsilon\right\}  =0,
\]
where $||a||$ is the $L_{\infty}$-norm of vector $a$.

To determine the fraction vector $S\left(  t\right)  $, we need to set up a
system of differential vector equations satisfied by the fraction measure
$S\left(  t\right)  $ by means of density-dependent jump Markov processes.
Consider the supermarket model with $n$ servers, and determine the expected
change in the number of queues with at least $k$ customers over a small time
period of length d$t$. The probability vector that during this time period,
any arriving customer joins a queue of size $k-1$ is given by%
\[
n\left[  \lambda S_{k-1}^{\odot d}\left(  n,t\right)  -\lambda S_{k}^{\odot
d}\left(  n,t\right)  \right]  \text{d}t.
\]
Similarly, the probability vector that a customer leaves a server queued by
$k$ customers is given by%
\[
n\left[  S_{k}\left(  n,t\right)  T+\sum_{l=1}^{\infty}b_{l}S_{k+l}\left(
n,t\right)  T^{0}\alpha\right]  \text{d}t.
\]
Therefore, we obtain%
\begin{align*}
\text{d}E\left[  n_{k}\left(  n,t\right)  \right]  =  &  n\left[  \lambda
S_{k-1}^{\odot d}\left(  n,t\right)  -\lambda S_{k}^{\odot d}\left(
n,t\right)  \right]  \text{d}t\\
&  +n\left[  S_{k}\left(  n,t\right)  T+\sum_{l=1}^{\infty}b_{l}S_{k+l}\left(
n,t\right)  T^{0}\alpha\right]  \text{d}t.
\end{align*}
This leads to%
\begin{equation}
\frac{\text{d}S_{k}\left(  n,t\right)  }{\text{d}t}=\lambda S_{k-1}^{\odot
d}\left(  n,t\right)  -\lambda S_{k}^{\odot d}\left(  n,t\right)
+S_{k}\left(  n,t\right)  T+\sum_{l=1}^{\infty}b_{l}S_{k+l}\left(  n,t\right)
T^{0}\alpha. \label{EqE1-1}%
\end{equation}
Taking $n\rightarrow\infty$ in both sides of Equation (\ref{EqE1-1}), we have%
\begin{equation}
\frac{\text{d}S_{k}\left(  t\right)  }{\text{d}t}=\lambda S_{k-1}^{\odot
d}\left(  t\right)  -\lambda S_{k}^{\odot d}\left(  t\right)  +S_{k}\left(
t\right)  T+\sum_{l=1}^{\infty}b_{l}S_{k+l}\left(  n,t\right)  T^{0}\alpha.
\label{EqE1-2}%
\end{equation}

Using a similar analysis to Equation (\ref{EqE1-2}), we obtain a system of
differential vector equations for the fraction vector $S\left(  t\right)
=\left(  S_{0}\left(  t\right)  ,S_{1}\left(  t\right)  ,S_{2}\left(
t\right)  ,\ldots\right)  $ as follows:%
\begin{equation}
S_{0}\left(  t\right)  =1, \label{EqE1-3}%
\end{equation}%
\begin{equation}
\frac{\mathtt{d}}{\text{d}t}S_{0}\left(  t\right)  =-\lambda S_{0}^{d}\left(
t\right)  +\sum_{l=1}^{\infty}S_{l}\left(  t\right)  T^{0}\sum_{k=l}^{\infty
}b_{k}, \label{EqE1-4}%
\end{equation}%
\begin{equation}
\frac{\mathtt{d}}{\text{d}t}S_{1}\left(  t\right)  =\lambda\alpha S_{0}%
^{d}\left(  t\right)  -\lambda S_{1}^{\odot d}\left(  t\right)  +S_{1}\left(
t\right)  T+\sum_{l=1}^{\infty}b_{l}S_{1+l}\left(  t\right)  T^{0}\alpha,
\label{EqE1-5}%
\end{equation}
and for $k\geq2$,%
\begin{equation}
\frac{\mathtt{d}}{\text{d}t}S_{k}\left(  t\right)  =\lambda S_{k-1}^{\odot
d}\left(  t\right)  -\lambda S_{k}^{\odot d}\left(  t\right)  +S_{k}\left(
t\right)  T+\sum_{l=1}^{\infty}b_{l}S_{k+l}\left(  t\right)  T^{0}\alpha.
\label{EqE1-6}%
\end{equation}

If the row vector $\pi=\left(  \pi_{0},\pi_{1},\pi_{2},\ldots\right)  $ is a
fixed point of the fraction vector $S\left(  t\right)  $, then the fixed point
$\pi$ satisfies the following system of nonlinear equations%
\begin{equation}
\pi_{0}=1 \label{EqE1-7}%
\end{equation}%
\begin{equation}
-\lambda\pi_{0}^{d}+\sum_{l=1}^{\infty}\pi_{l}T^{0}\sum_{k=l}^{\infty}b_{k}=0,
\label{EqE1-8}%
\end{equation}%
\begin{equation}
\lambda\alpha\pi_{0}^{d}-\lambda\pi_{1}^{\odot d}+\pi_{1}T+\sum_{l=1}^{\infty
}b_{l}\pi_{1+l}T^{0}\alpha=0, \label{EqE1-9}%
\end{equation}
and for $k\geq2$,%
\begin{equation}
\lambda\pi_{k-1}^{\odot d}-\lambda\pi_{k}^{\odot d}+\pi_{k}T+\sum
_{l=1}^{\infty}b_{l}\pi_{k+l}T^{0}\alpha=0. \label{EqE1-10}%
\end{equation}

Let%
\[
Q_{\text{right}}=\left(
\begin{array}
[c]{cccccc}%
-\lambda & \lambda\alpha &  &  &  & \cdots\\
& -\lambda I & \lambda I &  &  & \cdots\\
&  & -\lambda I & \lambda I &  & \cdots\\
&  &  & -\lambda I & \lambda I & \cdots\\
&  &  &  & \ddots & \ddots
\end{array}
\right)
\]
and%
\[
Q_{\text{left}}=\left(
\begin{array}
[c]{cccccc}%
0 &  &  &  &  & \\
T^{0} & T &  &  &  & \\
T^{0}\sum\limits_{k=2}^{\infty}b_{k} & b_{1}T^{0}\alpha & T &  &  & \\
T^{0}\sum\limits_{k=3}^{\infty}b_{k} & b_{2}T^{0}\alpha & b_{1}T^{0}\alpha &
T &  & \\
T^{0}\sum\limits_{k=4}^{\infty}b_{k} & b_{3}T^{0}\alpha & b_{2}T^{0}\alpha &
b_{1}T^{0}\alpha & T & \\
\vdots & \vdots & \vdots & \vdots & \vdots & \ddots
\end{array}
\right)  .
\]
Then the system of differential vector equations for the fraction measure is
given by%
\[
S_{0}\left(  t\right)  =1,
\]
and%
\[
\frac{\text{d}}{\text{d}t}S\left(  t\right)  =S^{\odot d}\left(  t\right)
Q_{\text{right}}+S\left(  t\right)  Q_{\text{left}};
\]
and the system of nonlinear equations for the fixed point is given by%
\[
\pi_{0}=1,
\]
and%
\begin{equation}
\pi^{\odot d}Q_{\text{right}}+\pi Q_{\text{left}}=0. \label{EqE1-11}%
\end{equation}

In the remainder of this section, we provide an iterative algorithm for
computing the fixed point for the supermarket model of GI/M/1 type.
Specifically, the iterative algorithm indicates that the supermarket model of
GI/M/1 type is more difficult than the case of M/G/1 type.

Let%
\[
B=\left(
\begin{array}
[c]{cccc}%
-\lambda I & \lambda I &  & \cdots\\
& -\lambda I & \lambda I & \cdots\\
&  & -\lambda I & \cdots\\
&  &  & \ddots
\end{array}
\right)
\]
and%
\[
Q_{\text{service}}=\left(
\begin{array}
[c]{ccccc}%
T &  &  &  & \\
b_{1}T^{0}\alpha & T &  &  & \\
b_{2}T^{0}\alpha & b_{1}T^{0}\alpha & T &  & \\
b_{3}T^{0}\alpha & b_{2}T^{0}\alpha & b_{1}T^{0}\alpha & T & \\
\vdots & \vdots & \vdots & \vdots & \ddots
\end{array}
\right)
\]
Then it follows from (\ref{EqE1-11}) that%
\begin{equation}
\pi_{0}^{d}\left(  \lambda\alpha,0,0,0,\ldots\right)  +\pi_{\mathcal{L}%
}^{\odot d}B+\pi_{\mathcal{L}}Q_{\text{Service}}=0, \label{EqE1-12}%
\end{equation}
where $\pi_{\mathcal{L}}=\left(  \pi_{1},\pi_{2},\pi_{3},\ldots\right)  $.
Note that%
\[
\left(  -B\right)  ^{-1}=\left(
\begin{array}
[c]{cccc}%
\frac{1}{\lambda}I & \frac{1}{\lambda}I & \frac{1}{\lambda}I & \cdots\\
& \frac{1}{\lambda}I & \frac{1}{\lambda}I & \cdots\\
&  & \frac{1}{\lambda}I & \cdots\\
&  &  & \ddots
\end{array}
\right)  ,
\]
using $\pi_{0}=1$ we obtain%
\[
\pi_{0}^{d}\left(  \lambda\alpha,0,0,0,\ldots\right)  \left(  -B\right)
^{-1}=\left(  \alpha,\alpha,\alpha,\ldots\right)
\]
and%
\[
Q_{\text{service}}\left(  -B\right)  ^{-1}=\frac{1}{\lambda}\left(
\begin{array}
[c]{ccccc}%
T & T & T & T & \cdots\\
\left(  T^{0}\alpha\right)  b_{1} & T+\left(  T^{0}\alpha\right)  b_{1} &
T+\left(  T^{0}\alpha\right)  b_{1} & T+\left(  T^{0}\alpha\right)  b_{1} &
\cdots\\
\left(  T^{0}\alpha\right)  b_{2} & \left(  T^{0}\alpha\right)  \sum
\limits_{k=1}^{2}b_{k} & T+\left(  T^{0}\alpha\right)  \sum\limits_{k=1}%
^{2}b_{k} & T+\left(  T^{0}\alpha\right)  \sum\limits_{k=1}^{2}b_{k} &
\cdots\\
\left(  T^{0}\alpha\right)  b_{3} & \left(  T^{0}\alpha\right)  \sum
\limits_{k=2}^{3}b_{k} & \left(  T^{0}\alpha\right)  \sum\limits_{k=1}%
^{3}b_{k} & T+\left(  T^{0}\alpha\right)  \sum\limits_{k=1}^{3}b_{k} &
\cdots\\
\vdots & \vdots & \vdots & \vdots &
\end{array}
\right)  ,
\]
Thus it follows from (\ref{EqE1-12}) that%
\begin{align*}
\pi_{\mathcal{L}}^{\odot d}  &  =\pi_{0}^{d}\left(  \lambda\alpha
,0,0,0,\ldots\right)  \left(  -B\right)  ^{-1}+\pi_{\mathcal{L}}%
Q_{\text{service}}\left(  -B\right)  ^{-1}\\
&  =\left(  \alpha,\alpha,\alpha,\ldots\right)  +\pi_{\mathcal{L}%
}Q_{\text{service}}\left(  -B\right)  ^{-1}.
\end{align*}
This leads to that for $k\geq1$%
\begin{equation}
\lambda\pi_{k}^{d}=\lambda\alpha+\sum\limits_{l=1}^{k+1}\pi_{l}T+b_{1}%
\sum\limits_{l=2}^{k+2}\pi_{l}\left(  T^{0}\alpha\right)  +b_{2}%
\sum\limits_{l=3}^{k+3}\pi_{l}\left(  T^{0}\alpha\right)  +\cdots.
\label{EqE1-14}%
\end{equation}
To solve the system of nonlinear equations (\ref{EqE1-14}) for $k\geq1$, we
assume that the fixed point has a closed-form solution%
\[
\pi_{k}=r\left(  k\right)  \eta,
\]
where $\eta$ is the stationary probability vector of the Markov chain
$T+T^{0}\alpha$. It follows from (\ref{EqE1-14}) that%
\[
\lambda r^{d}\left(  k\right)  \eta^{\odot d}=\lambda\alpha+\sum
\limits_{l=1}^{k+1}r\left(  l\right)  \eta T+b_{1}\sum\limits_{l=2}%
^{k+2}r\left(  l\right)  \eta\left(  T^{0}\alpha\right)  +b_{2}\sum
\limits_{l=3}^{k+3}r\left(  l\right)  \eta\left(  T^{0}\alpha\right)
+\cdots.
\]
Taking $\theta=\eta^{\odot d}e$. Then $\theta\in\left(  0,1\right)  $. Noting
that $\alpha e=1,\eta Te=-\mu$ and $\eta T^{0}=\mu$, we obtain that for
$k\geq1$
\begin{equation}
\rho\theta r^{d}\left(  k\right)  =\rho-\sum\limits_{l=1}^{k+1}r\left(
l\right)  +b_{1}\sum\limits_{l=2}^{k+2}r\left(  l\right)  +b_{2}%
\sum\limits_{l=3}^{k+3}r\left(  l\right)  +\cdots. \label{EqGI-1}%
\end{equation}
This gives%
\begin{equation}
\rho\theta\left[  r^{d}\left(  k\right)  -r^{d}\left(  k+1\right)  \right]
=r\left(  k+2\right)  -\sum\limits_{l=1}^{\infty}b_{l}r\left(  k+2+l\right)  .
\label{EqGI-2}%
\end{equation}
Thus it follows from (\ref{EqGI-2}) that%
\begin{equation}
\rho\theta\left(  r^{d}\left(  1\right)  -r^{d}\left(  2\right)  ,r^{d}\left(
2\right)  -r^{d}\left(  3\right)  ,r^{d}\left(  3\right)  -r^{d}\left(
4\right)  ,\ldots\right)  =\left(  r\left(  3\right)  ,r\left(  4\right)
,r\left(  5\right)  ,\ldots\right)  C, \label{EqGI-3}%
\end{equation}
where%
\[
C=\left(
\begin{array}
[c]{ccccc}%
1 &  &  &  & \\
-b_{1} & 1 &  &  & \\
-b_{2} & -b_{1} & 1 &  & \\
-b_{3} & -b_{2} & -b_{1} & 1 & \\
\vdots & \vdots & \vdots & \vdots & \ddots
\end{array}
\right)  .
\]
Therefore, we have%
\[
C^{-1}=\left(
\begin{array}
[c]{cccccc}%
1 &  &  &  &  & \\
b_{1} & 1 &  &  &  & \\
b_{2}+b_{1}^{2} & b_{1} & 1 &  &  & \\
b_{3}+2b_{2}b_{1}+b_{1}^{3} & b_{2}+b_{1}^{2} & b_{1} & 1 &  & \\
b_{4}+2b_{3}b_{1}+3b_{2}b_{1}^{2}+b_{1}^{4} & b_{3}+2b_{2}b_{1}+b_{1}^{3} &
b_{2}+b_{1}^{2} & b_{1} & 1 & \\
\vdots & \vdots & \vdots & \vdots & \vdots & \ddots
\end{array}
\right)  ,
\]
and the norm $||\cdot||_{\infty}$ of the matrix $C^{-1}$ is given by
\[
||C^{-1}||=\sup_{i\geq j\geq1}\left\{  |\zeta_{i,j}|\right\}  \leq\sum
_{k=1}^{\infty}b_{k}=1,
\]
where $\zeta_{i,j}$ is the $\left(  i,j\right)  $th entry of the matrix
$C^{-1}$ for $0\leq j\leq i$. Note that the norm $||C^{-1}||\leq1$ is useful
for our following iterative algorithm designed by the matrix $\rho\theta
C^{-1}$ with $||\rho\theta C^{-1}||=\rho\theta<1$.

Let
\[
X=\left(  r^{d}\left(  1\right)  -r^{d}\left(  2\right)  ,r^{d}\left(
2\right)  -r^{d}\left(  3\right)  ,r^{d}\left(  3\right)  -r^{d}\left(
4\right)  ,\ldots\right)
\]
and%
\[
Y=\left(  r\left(  3\right)  ,r\left(  4\right)  ,r\left(  5\right)
,\ldots\right)  .
\]
Then%
\[
X=\left(  r^{d}\left(  1\right)  ,r^{d}\left(  2\right)  ,Y^{\odot d}\right)
-\left(  r^{d}\left(  2\right)  ,Y^{\odot d}\right)
\]
and it follows from (\ref{EqGI-3}) that%

\begin{align}
Y  &  =X\left(  \rho\theta C^{-1}\right) \nonumber\\
&  =\left(  r^{d}\left(  1\right)  ,r^{d}\left(  2\right)  ,Y^{\odot
d}\right)  \left(  \rho\theta C^{-1}\right)  -\left(  r^{d}\left(  2\right)
,Y^{\odot d}\right)  \left(  \rho\theta C^{-1}\right)  . \label{EqGI-4}%
\end{align}
It follows from (\ref{EqGI-1}) that%
\begin{equation}
r\left(  1\right)  =\rho-\rho\theta r^{d}\left(  1\right)  -r\left(  2\right)
\left(  1-b_{1}\right)  +Y\left(  b_{1}+b_{2},b_{2}+b_{3},b_{3}+b_{4}%
,\ldots\right)  ^{T} \label{EqGI-5}%
\end{equation}
and%
\begin{equation}
r\left(  2\right)  =\rho-r\left(  1\right)  -\left[  b_{1}r\left(  2\right)
+\rho\theta r^{d}\left(  2\right)  \right]  +Y\left(  b_{1}+b_{2}%
-1,b_{1}+b_{2}+b_{3},b_{2}+b_{3}+b_{4},\ldots\right)  ^{T} \label{EqGI-6}%
\end{equation}

Now, we use Equations (\ref{EqGI-4}) to (\ref{EqGI-6}) to provide an iterative
algorithm for computing the fixed point $\pi_{k}=r\left(  k\right)  \eta$ for
$k\geq1$. To that end, we write%
\begin{equation}
Y_{N}=\left(  r_{N}\left(  3\right)  ,r_{N}\left(  4\right)  ,r_{N}\left(
5\right)  ,\ldots\right)  \label{EqGI-6-1}%
\end{equation}
and%
\begin{equation}
R_{N}=\left(  r_{N}\left(  1\right)  ,r_{N}\left(  2\right)  ,r_{N}\left(
3\right)  ,\ldots\right)  =\left(  r_{N}\left(  1\right)  ,r_{N}\left(
2\right)  ,Y_{N}\right)  . \label{EqGI-6-2}%
\end{equation}
Let%
\begin{align}
r_{N+1}\left(  1\right)  =  &  \rho-\rho\theta r_{N}^{d}\left(  1\right)
-r_{N}\left(  2\right)  \left(  1-b_{1}\right) \nonumber\\
&  +Y_{N}\left(  b_{1}+b_{2},b_{2}+b_{3},b_{3}+b_{4},\ldots\right)  ^{T},
\label{EqGI-7}%
\end{align}%
\begin{align}
r_{N+1}\left(  2\right)  =  &  \rho-r_{N}\left(  1\right)  -\left[  b_{1}%
r_{N}\left(  2\right)  +\rho\theta r_{N}^{d}\left(  2\right)  \right]
\nonumber\\
&  +Y_{N}\left(  b_{1}+b_{2}-1,b_{1}+b_{2}+b_{3},b_{2}+b_{3}+b_{4}%
,\ldots\right)  ^{T} \label{EqGI-8}%
\end{align}
and%
\begin{equation}
Y_{N+1}=\left(  r_{N}^{d}\left(  1\right)  ,r_{N}^{d}\left(  2\right)
,Y_{N}^{\odot d}\right)  \left(  \rho\theta C^{-1}\right)  -\left(  r_{N}%
^{d}\left(  2\right)  ,Y_{N}^{\odot d}\right)  \left(  \rho\theta
C^{-1}\right)  . \label{EqGI-9}%
\end{equation}

Based on the iterative relations given in (\ref{EqGI-7}) to (\ref{EqGI-9}), we
provide an iterative algorithm for computing the vector $R=\left(  r\left(
1\right)  ,r\left(  2\right)  ,r\left(  3\right)  ,\ldots\right)  $. This
gives the fixed point $\pi=\left(  1,r\left(  1\right)  \eta,r\left(
2\right)  \eta,r\left(  3\right)  \eta,\ldots\right)  $.

\textbf{An Iterative Algorithm: }Computation of the Fixed Point

\texttt{Input:} $\ \ \lambda,\left(  \alpha,T\right)  ,\left\{  b_{k}\right\}
$ and $d$.

\texttt{Output:} \ $R=\left(  r\left(  1\right)  ,r\left(  2\right)  ,r\left(
3\right)  ,\ldots\right)  $ and $\pi=\left(  1,r\left(  1\right)
\eta,r\left(  2\right)  \eta,r\left(  3\right)  \eta,\ldots\right)  $.

\texttt{Computational Steps:}

\emph{Step one:} Taking the initial value $R_{0}=0$, that is, $r_{0}\left(
1\right)  =0,r_{0}\left(  2\right)  =0,Y_{0}=0$.

\emph{Step two:} Computing $R_{1}=\left(  r_{1}\left(  1\right)  ,r_{1}\left(
2\right)  ,Y_{1}\right)  $ through%
\[%
\begin{array}
[c]{ll}%
r_{1}\left(  1\right)  =\rho, & \text{ \ }\leftarrow\text{(\ref{EqGI-7})}\\
r_{1}\left(  2\right)  =\rho, & \text{ \ }\leftarrow\text{(\ref{EqGI-8})}\\
Y_{1}=\left(  \rho^{d},\rho^{d},0\right)  \left(  \rho\theta C^{-1}\right)
-\left(  \rho^{d},0\right)  \left(  \rho\theta C^{-1}\right)  , & \text{
\ }\leftarrow\text{(\ref{EqGI-9})}%
\end{array}
\]

\emph{Step three:} If $R_{N}$ is known, computing $R_{N+1}=\left(
r_{N+1}\left(  1\right)  ,r_{N+1}\left(  2\right)  ,Y_{N+1}\right)  $ through%
\[%
\begin{array}
[c]{ll}%
\begin{array}
[c]{l}%
r_{N+1}\left(  1\right)  =\rho-\rho\theta r_{N}^{d}\left(  1\right)
-r_{N}\left(  2\right)  \left(  1-b_{1}\right) \\
\text{ \ }+Y_{N}\left(  b_{1}+b_{2},b_{2}+b_{3},b_{3}+b_{4},\ldots\right)
^{T},
\end{array}
& \text{ \ }\leftarrow\text{(\ref{EqGI-7})}\\%
\begin{array}
[c]{l}%
r_{N+1}\left(  2\right)  =\rho-r_{N}\left(  1\right)  -\left[  b_{1}%
r_{N}\left(  2\right)  +\rho\theta r_{N}^{d}\left(  2\right)  \right] \\
\text{ \ }+Y_{N}\left(  b_{1}+b_{2}-1,b_{1}+b_{2}+b_{3},b_{2}+b_{3}%
+b_{4},\ldots\right)  ^{T},
\end{array}
& \text{ \ }\leftarrow\text{(\ref{EqGI-8})}\\
Y_{N+1}=\left(  r_{N}^{d}\left(  1\right)  ,r_{N}^{d}\left(  2\right)
,Y_{N}^{\odot d}\right)  \left(  \rho\theta C^{-1}\right)  -\left(  r_{N}%
^{d}\left(  2\right)  ,Y_{N}^{\odot d}\right)  \left(  \rho\theta
C^{-1}\right)  . & \text{ \ }\leftarrow\text{(\ref{EqGI-9})}%
\end{array}
\]

\emph{Step four:} For a sufficiently small $\varepsilon>0$, if there exists
Step $K$ such that $||R_{K+1}-R_{K}||<\varepsilon$, then our computation is
end in this step; otherwise we go to Step three for continuous computations.

\emph{Step five:} When our computation is over at Step $K$, computing%
\[
\pi=\left(  1,r_{K}\left(  1\right)  \eta,r_{K}\left(  2\right)  \eta
,r_{K}\left(  3\right)  \eta,\ldots\right)
\]
as an approximate fixed point under an error $\varepsilon>0$.

In what follows we analyze two numerical examples by means of the above
iterative algorithm.

In the first example, we take%
\[
\lambda=1,d=2,\alpha=\left(  1/2,1/2\right)  ,
\]%
\[
T\left(  1\right)  =\left(
\begin{array}
[c]{cc}%
-4 & 3\\
2 & -7
\end{array}
\right)  ,T\left(  2\right)  =\left(
\begin{array}
[c]{cc}%
-5 & 3\\
2 & -7
\end{array}
\right)  ,T\left(  3\right)  =\left(
\begin{array}
[c]{cc}%
-4 & 4\\
2 & -7
\end{array}
\right)  ,
\]
Table \ref{table: 1} illustrates how the super-exponential solution ($\pi_{1}$
to $\pi_{5}$) depends on the matrices $T\left(  1\right)  $, $T\left(
2\right)  $ and $T\left(  3\right)  $, respectively.

\begin{table}[tbh]
\caption{The super-exponential solution depends on the matrix $T$}%
\label{table: 1}%
\centering      {\scriptsize
\begin{tabular}
[c]{|c|c|c|c|}\hline
& $T(1)$ & $T(2)$ & $T(3)$\\\hline\hline
$\pi_{1}$ & (0.2045,\;0.1591) & (0.1410,\;0.1026) & (0.3125,\; 0.2500)\\\hline
$\pi_{2}$ & (0.0137,\;0.0107) & (0.0043,\;0.0031) & (0.0500,\;0.0400)\\\hline
$\pi_{3}$ & (6.193e-05,\;4.817e-05) & (3.965e-06,\;2.884e-06) & (0.0013
,\;0.0010)\\\hline
$\pi_{4}$ & (1.259e-09,\;9.793e-10) & (3.390e-12,\;2.465e-12) &
(8.446e-07,\;6.757e-07)\\\hline
$\pi_{5}$ & (5.204e-19,\;4.048e-19) & (2.478e-24,\;1.802e-24) & (3.656e-13,\;
2.925e-13)\\\hline
\end{tabular}
}\end{table}

In the second example, we take%
\[
\lambda=1,d=5,\alpha\left(  1\right)  =\left(  1/3,1/3,1/3\right)
,\alpha\left(  2\right)  =\left(  1/12,7/12,1/3\right)  ,
\]%
\[
T=\left(
\begin{array}
[c]{ccc}%
-10 & 2 & 4\\
3 & -7 & 4\\
0 & 2 & -5
\end{array}
\right)  ,
\]
Table \ref{table: 2} shows how the super-exponential solution ($\pi_{1}$ to
$\pi_{4}$) depends on the vectors $\alpha\left(  1\right)  $ and
$\alpha\left(  2\right)  $, respectively.

\begin{table}[tbh]
\caption{The super-exponential solution depends on the vectors $\alpha$}%
\label{table: 2}%
\centering      {\scriptsize
\begin{tabular}
[c]{|c|c|c|}\hline
& $\alpha=(\frac{1}{3},\frac{1}{3},\frac{1}{3})$ & $\alpha=(\frac{1}%
{12},\frac{7}{12},\frac{1}{3})$\\\hline\hline
$\pi_{1}$ & (0.0741,\;0.1358 ,\;0.2346) & (0.0602,\;0.1728,\;0.2531)\\\hline
$\pi_{2}$ & (5.619e-05,\;1.030e-05,\; 1.779e-04 ) &
(7.182e-05,\;2.063e-04,\;3.020e-04)\\\hline
$\pi_{3}$ & (1.411e-20,\;2.587e-20,\;4.469e-20) &
(1.739e-19,\;4.993e-19,\;7.311e-19)\\\hline
$\pi_{4}$ & (1.410e-98,\;2.586e-98,\;4.466e-98) &
(1.444e-92,\;4.148e-92,\;6.074e-92)\\\hline
\end{tabular}
}\end{table}

\section{Supermarket Models with Multiple Choices}

In this section, we consider two supermarket models with multiple choices: The
first one is one mobile server with multiple waiting lines under the service
discipline of joint-shortest queue and serve-longest queue, and the second one
is a supermarket model with multiple classes of Poisson arrivals, each of
which has a choice number. Our main purpose is to organize the system of
nonlinear equations for the fixed point under multiple choice numbers, and to
be able to obtain super-exponential solution to the fixed points for the two
supermarket models.

\subsection{One mobile server with multiple waiting lines}

The supermarket model is structured as one mobile server with multiple waiting
lines, where the Poisson arrivals joint a waiting line with the shortest queue
and the mobile server enters a waiting line with the longest queue for his
service woks. Such a system is depicted in Figure 3 for an illustration. For
one mobile server with $n$ waiting lines, customers arrive at this system as a
Poisson process with arrival rate $n\lambda$, and all customers are served by
one mobile server with service rate $n\mu$. Each arriving customer chooses
$d\geq1$ waiting lines independently and uniformly at random from the $n$
waiting lines, and waits for service at a waiting line which currently
contains the fewest number of customers. If there is a tie, waiting lines with
the fewest number of customers will be chosen by the arriving customer
randomly. The mobile server chooses $f\geq1$ waiting lines independently and
uniformly at random from the $n$ waiting lines, and enters a waiting line
which currently contains the most number of customers. If there is a tie,
waiting lines with the most number of customers will be chosen be the server
randomly. All customers in every waiting line will be served in the FCFS
manner. We assume that all random variables defined above are independent of
each other, and that the system is operating in the stable region
$\rho=\lambda/\mu<1$. Clearly, $d$ and $f$ are input choice number and output
choice number in this supermarket model, respectively.

\begin{figure}[ptbh]
\centering                    \includegraphics[width=10cm]{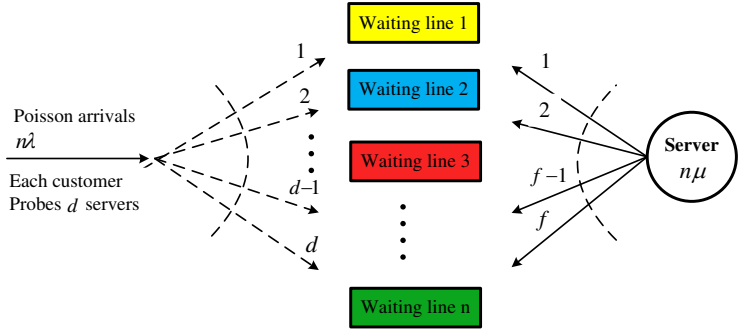}
\caption{A supermarket model with input and output choices}%
\label{figure: model-3}%
\end{figure}

It is clear that the stochastic environment of this supermarket model is a
positive recurrent birth-death process with an irreducible infinitesimal
generator $Q=Q_{\text{left}}+Q_{\text{right}}$, where%
\[
Q_{\text{left}}=\left(
\begin{array}
[c]{ccccc}%
0 &  &  &  & \\
\mu & -\mu &  &  & \\
& \mu & -\mu &  & \\
&  & \mu & -\mu & \\
&  &  & \ddots & \ddots
\end{array}
\right)
\]
and%
\[
Q_{\text{right}}=\left(
\begin{array}
[c]{ccccc}%
-\lambda & \lambda &  &  & \\
& -\lambda & \lambda &  & \\
&  & -\lambda & \lambda & \\
&  &  & \ddots & \ddots
\end{array}
\right)  .
\]
Similar derivation to those given in Section 3 or 4, we obtain that the fixed
point satisfies the system of nonlinear equations%
\[
\pi_{0}=1
\]
and%
\begin{equation}
\pi^{\odot f}Q_{\text{left}}+\pi^{\odot d}Q_{\text{right}}=0. \label{EqE1-0}%
\end{equation}

Let%
\[
Q=\left(
\begin{array}
[c]{cc}%
Q_{0,0} & U\\
V & Q^{\left(  \mathcal{L}\right)  }%
\end{array}
\right)  ,
\]
where%
\[
Q_{0,0}=-\lambda,U=\left(  \lambda,0,0,\ldots\right)  ,V=\left(
\mu,0,0,\ldots\right)  ^{T},
\]%
\[
Q_{\text{arrival}}^{\left(  \mathcal{L}\right)  }=\left(
\begin{array}
[c]{ccccc}%
-\lambda & \lambda &  &  & \\
& -\lambda & \lambda &  & \\
&  & -\lambda & \lambda & \\
&  &  & \ddots & \ddots
\end{array}
\right)
\]
and%
\[
Q_{\text{service}}^{\left(  \mathcal{L}\right)  }=\left(
\begin{array}
[c]{cccc}%
-\mu &  &  & \\
\mu & -\mu &  & \\
& \mu & -\mu & \\
&  & \ddots & \ddots
\end{array}
\right)  .
\]
It follows from (\ref{EqE1-0}) that%
\begin{equation}
\pi_{0}=1, \label{EquE1-1}%
\end{equation}%
\begin{equation}
-\lambda\pi_{0}^{d}+\mu\pi_{1}^{f}=0 \label{EquE1-2}%
\end{equation}
and%
\begin{equation}
\pi_{0}^{d}U+\pi_{\mathcal{L}}^{\odot d}Q_{\text{arrival}}^{\left(
\mathcal{L}\right)  }+\pi_{\mathcal{L}}^{\odot f}Q_{\text{service}}^{\left(
\mathcal{L}\right)  }=0. \label{EquE1-3}%
\end{equation}
It follows from (\ref{EquE1-1}) and (\ref{EquE1-2}) that%
\[
\pi_{1}=\rho^{\frac{1}{f}}.
\]
Note that%
\[
\left[  -Q_{\text{service}}^{\left(  \mathcal{L}\right)  }\right]
^{-1}=\left(
\begin{array}
[c]{cccc}%
\frac{1}{\mu} &  &  & \\
\frac{1}{\mu} & \frac{1}{\mu} &  & \\
\frac{1}{\mu} & \frac{1}{\mu} & \frac{1}{\mu} & \\
\vdots & \vdots & \vdots & \ddots
\end{array}
\right)  ,
\]
it follows from (\ref{EquE1-3}) that for $k\geq2$%
\[
\pi_{k}^{f}=\pi_{k-1}^{d}\rho.
\]
This leads to%
\begin{equation}
\pi_{k}=\rho^{\frac{\sum\limits_{i=0}^{k-1}d^{i}f^{k-1-i}}{f^{k}}}%
=\rho^{\frac{1}{f}\sum\limits_{i=0}^{k-1}\left(  \frac{d}{f}\right)  ^{i}}.
\label{EquE1-4}%
\end{equation}
Specifically, when $d\neq f$, we have%
\[
\pi_{k}=\rho^{\frac{\left(  \frac{d}{f}\right)  ^{k}-1}{d-f}}.
\]

\begin{Rem}
Equation (\ref{EquE1-4}) indicates different influence of the input and output
choice numbers $d$ and $f$ on the fixed point $\pi$. If $d>f$, then the fixed
point $\pi$ decreases doubly exponentially; and if $d=f$, then $\pi_{k}%
=\rho^{\frac{k}{f}}$ which is geometric. However, it is very interesting for
the case with $d<f$. In this case, $\lim_{k\rightarrow\infty}\pi_{k}%
=\rho^{\frac{1}{f-d}}$, which illustrates that the fraction of waiting lines
with infinite customers has a positive lower bound $\rho^{\frac{1}{f-d}}>0$.
This shows that if $\rho<1$ and $d<f$, this supermarket model is transient.
\end{Rem}

\subsection{A supermarket model with multiple input choices}

Now, we analyze a supermarket model with multiple input choices. There are $m$
types of different customers who arrive at a queueing system of $n>1$ servers
for receiving their required service. Arrivals of customers of $i$th type are
a Poisson process with arrival rate $n\lambda_{i}$ for $\lambda_{i}>0$, and
the service times at each server are exponential with service rate $\mu>0$.
Note that different types of customers have the same service time. Each
arriving customer of $i$th type chooses $d_{i}\geq1$ servers independently and
uniformly at random from the $n$ servers, and waits for service at the server
which currently contains the fewest number of customers. If there is a tie,
servers with the fewest number of customers will be chosen randomly. All
customers in every server will be served in the FCFS manner. We assume that
all random variables defined above are independent of each other, and that the
system is operating in the stable region $\rho=\sum_{i=1}^{m}\rho_{i}<1$,
where $\rho_{i}=\lambda_{i}/\mu$. Clearly, $d_{1},d_{2},\ldots,d_{m}$ are
multiple input choice numbers in this supermarket model.

Let%
\[
Q_{\text{right}}\left(  i\right)  =\left(
\begin{array}
[c]{ccccc}%
-\lambda_{i} & \lambda_{i} &  &  & \\
& -\lambda_{i} & \lambda_{i} &  & \\
&  & -\lambda_{i} & \lambda_{i} & \\
&  &  & \ddots & \ddots
\end{array}
\right)
\]
and%
\[
Q_{\text{left}}=\left(
\begin{array}
[c]{cccc}%
0 &  &  & \\
\mu & -\mu &  & \\
& \mu & -\mu & \\
&  & \ddots & \ddots
\end{array}
\right)  .
\]
Obviously, the stochastic environment of this supermarket model is a positive
recurrent birth-death process with an irreducible infinitesimal generator
$Q=Q_{\text{left}}+\sum_{i=1}^{m}Q_{\text{right}}\left(  i\right)  $. Similar
derivation to those given in Section 3 or 4, we obtain that the fixed point
satisfies the system of nonlinear equations%
\begin{equation}
\pi_{0}=1 \label{EqE3-1}%
\end{equation}
and%
\begin{equation}
\pi Q_{\text{left}}+\sum_{i=1}^{m}\pi^{\odot d_{i}}Q_{\text{right}}\left(
i\right)  =0. \label{EqE3-2}%
\end{equation}
Let%
\[
\pi=\left(  \pi_{0},\pi_{\mathcal{L}}\right)  ,
\]%
\[
U_{i}=\left(  \lambda_{i},0,0,\ldots\right)  ,
\]%
\[
Q_{\text{arrival}}^{\left(  \mathcal{L}\right)  }\left(  i\right)  =\left(
\begin{array}
[c]{ccccc}%
-\lambda_{i} & \lambda_{i} &  &  & \\
& -\lambda_{i} & \lambda_{i} &  & \\
&  & -\lambda_{i} & \lambda_{i} & \\
&  &  & \ddots & \ddots
\end{array}
\right)
\]
and%
\[
Q_{\text{service}}^{\left(  \mathcal{L}\right)  }=\left(
\begin{array}
[c]{cccc}%
-\mu &  &  & \\
\mu & -\mu &  & \\
& \mu & -\mu & \\
&  & \ddots & \ddots
\end{array}
\right)  .
\]
Therefore, the system of nonlinear equations (\ref{EqE3-1}) and (\ref{EqE3-2})
is written as%
\begin{equation}
\pi_{0}=1, \label{EqE3-4}%
\end{equation}%
\begin{equation}
-\sum_{i=1}^{m}\lambda_{i}\pi_{0}^{d_{i}}+\mu\pi_{1}=0, \label{EqE3-5}%
\end{equation}%
\begin{equation}
\sum_{i=1}^{m}\pi_{0}^{d_{i}}\left(  \lambda_{i},0,0,\ldots\right)
+\sum_{i=1}^{m}\pi_{\mathcal{L}}^{\odot d_{i}}Q_{\text{arrival}}^{\left(
\mathcal{L}\right)  }\left(  i\right)  +\pi_{\mathcal{L}}Q_{\text{service}%
}^{\left(  \mathcal{L}\right)  }=0. \label{EqE3-6}%
\end{equation}
It follows from (\ref{EqE3-4}) and (\ref{EqE3-5}) that%
\[
\pi_{1}=\sum_{i=1}^{m}\rho_{i}=\rho,
\]
and from (\ref{EqE3-6}) that%
\[
\pi_{\mathcal{L}}=\sum_{i=1}^{m}\pi_{0}^{d_{i}}\left(  \lambda_{i}%
,0,0,\ldots\right)  \left[  -Q_{\text{service}}^{\left(  \mathcal{L}\right)
}\right]  ^{-1}+\sum_{i=1}^{m}\pi_{\mathcal{L}}^{\odot d_{i}}Q_{\text{arrival}%
}^{\left(  \mathcal{L}\right)  }\left(  i\right)  \left[  -Q_{\text{service}%
}^{\left(  \mathcal{L}\right)  }\right]  ^{-1}.
\]
This leads to that for $k\geq2$%
\[
\pi_{k}=\sum_{i=1}^{m}\pi_{k-1}^{d_{i}}\rho_{i}.
\]

Let $\delta_{1}=\rho$ and $\delta_{k}=\sum_{i=1}^{m}\delta_{k-1}^{d_{i}}%
\rho_{i}$ for $k\geq2$. Then the fixed point has a super-exponential solution%
\[
\pi_{0}=1
\]
and for $k\geq1$%
\[
\pi_{k}=\delta_{k}.
\]

\section*{Acknowledgements}

Q.L. Li was supported by the National Science Foundation of China under grant
No. 10871114.

\end{document}